\newcommand{\xdeleted}[1]{}

\documentclass[astronomy,article,accept,pdftex,moreauthors]{Definitions/mdpi}

\usepackage[normalem]{ulem}

\firstpage{1} 
\pubvolume{1}
\issuenum{1}
\articlenumber{0}
\pubyear{2026}
\copyrightyear{2026}
\externaleditor{Firstname Lastname} % More than 1 editor, please add `` and '' before the last editor name
\datereceived{6 December 2025} 
\daterevised{21 January 2026} % Comment out if no revised date
\dateaccepted{30 January 2026} 
\datepublished{ } 
\hreflink{https://doi.org/}
\Title{A Small Patch Hypothesis in Cosmology}
\Author{Meir Shimon~$^{1,2}$\orcidA{}
} 

\AuthorNames{Meir Shimon}

\address{%
$^{1}$ \quad School of Physics and Astronomy, Tel Aviv University,
Tel Aviv 69978, Israel; meirs@tauex.tau.ac.il\\
$^{2}$ \quad Afeka Tel-Aviv Academic College of Engineering, Tel Aviv 69107, Israel}

\abstract{If our observable Universe is only a tiny region of a vastly larger and conformally older spacetime, then the usual formulations of the classical flatness and horizon problems of the Hot Big Bang can be reinterpreted as artifacts manifesting an observational selection effect; we occupy a small causal domain of a much larger causally-connected and possibly non-flat spacetime. 
A sufficiently large positive cosmological constant, $\Lambda$, sets the future asymptotic horizon scale of the observable Universe, $\sim$$\Lambda^{-1/2}$, thereby implying that the observable Universe may simply be a minute patch of a far larger pre-existing one, hereafter a Small Patch Hypothesis.
Importantly, this observational bound is purely geometric; regardless of when the Universe is observed, the maximum accessible scale is finite and fixed by $\Lambda$, independent of inflationary dynamics, anthropic arguments, or assumptions about the global hosting spacetime.
The externally possibly frozen past-eternal state implied by a pre-existing, causally connected spacetime motivates, but does not strictly require, viewing the perturbation field as being in (or arbitrarily close to) a coarse-grained maximum-entropy --- equilibrium --- configuration. 
Conditionalizing only on fixed mean and variance, a Gaussian distribution uniquely emerges, while the absence of entropy gradients corresponds to adiabaticity. In this work these features are therefore treated as plausible maximum-ignorance priors for super-horizon perturbations, rather than as rigorously derived consequences of a fully developed microscopic notion of gravitational entropy.
In this sense, inflation becomes one viable realization of the proposed Small Patch Hypothesis. Here, one particular non-inflationary alternative is considered {for illustrative purposes} in which a primordial spectrum $P_{\zeta}(k)$ of the gauge-invariant perturbation $\zeta$ that pre-dates the Big Bang grows logarithmically toward large scales, $k\rightarrow 0$, and in fact diverges at some finite $k_{c}$.
If $k_{c}\ll \Lambda^{-1/2}$, then our local cosmic 
patch probes only the regime where $\zeta\ll 1$ and appears exceptionally smooth. Over the comparatively narrow observable window, this $P_{\zeta}(k)$ mimics a slightly red-tilted, inflation-like spectrum. 
Rather than introducing high-energy new fields, this perspective frames large-scale homogeneity, isotropy, Gaussianity, adiabaticity, and the observed thermodynamic Arrow of Time as possible consequences of restricted observational access to a much larger Universe in equilibrium, 
rather than signatures of a unique early-Universe mechanism. Current observations cannot distinguish this logarithmically running spectrum from the standard power-law one, but future probes --- for example high-resolution 21-cm measurements of the Dark Ages --- may be able to falsify it.}


\begin{document}

\section{Introduction}\label{sec1}

A few puzzling properties of the Universe --- that include, among others, the flatness~\cite{Dicke1970} and horizon~\cite{Rindler1956,Misner1969} problems --- are naturally explained in the standard cosmological model by a brief phase of inflationary expansion in the very early Universe, e.g.,~\cite{Guth1981,Linde1982,Starobinsky1982}. Not only that inflation dynamically explains these features of the model away, but it also provides a quantum origin for the observed slightly red-tilted primordial power spectrum, e.g.,~\cite{MukhanovChibisov1981,Sasaki1983}. Inflation allegedly took place at very high energies, $\sim$$10^{15}$ GeV or somewhat lower.
As will be argued in the present work the flatness and horizon problems, as well as the slightly red-tilted power spectrum, seem to 
be consistent with an alternative geometrical (rather than dynamical) low-energy (potentially) classical phenomenon. This alternative viewpoint will be referred to in this work as the {\it Small Patch Hypothesis}, in which the observable Universe inhabits a tiny fractional volume of a vastly larger spacetime (for definiteness, one may keep in mind a globally Friedmann-Robertson-Walker (FRW) or asymptotically de Sitter parent solution with curvature radius and horizon scale much larger than the present Hubble radius, without attempting here to derive that global solution from more fundamental physics).

The discovery of the accelerated expansion of the Universe,~\cite{Riess1998,Perlmutter1999}, came as a surprise since it was clear from the outset that the naive estimate of vacuum energy density $\rho_{vac}\sim m_{pl}^{4}$~\cite{Weinberg1989}, where $m_{pl}$ is the Planck mass, is outright rejected by the very existence of a few billion years old Universe, and it was expected that $\rho_{vac}=0$ precisely by virtue of some yet-to-be-discovered symmetry principle. 
Since the observationally inferred value of $\rho_{vac}$ is unlikely to come from any sector of the standard model of particle physics, various estimated upper bounds were proposed based on anthropic reasoning, e.g.,~\cite{Weinberg1987,Efstathiou1995,Martel1998}.  

From the perspective of the present work, there should also be a lower limit on the cosmological constant, otherwise a `typical' observer might find himself observing a very inhomogeneous Universe on sufficiently large scales, undermining the underlying Cosmological Principle. Put alternatively, for the latter to 
apply to the observable Universe the cosmological constant, $\Lambda$, 
has to be sufficiently large for a given observed tilt 
of the matter power spectrum as will be argued below.
This consideration alone implies that a positive cosmological constant is likely to be in play and at just about the observed level. Similar conclusions, albeit from entirely different perspectives, have been reached at in, e.g.,~\cite{Shimon2024,AfshordiMagueijo2022}. In the context of the Small Patch Hypothesis, $\Lambda$ sets the observational boundary that regulates which Fourier modes of the primordial spectrum are accessible within the specific logarithmically running ansatz for $P_\zeta(k)$ developed in Section~\ref{sec4}. It should be stressed already at this point that the Small Patch Hypothesis is perfectly consistent with the standard slightly red-tilted power spectrum and the reason that a logarithmically running replacement is even considered in the present work is to demonstrate that even if the hosting Universe becomes wildly inhomogeneous on very large (yet finite) ultra-Hubble scales there is always a sufficiently large value of $\Lambda$ that would guarantee that the observable Universe is smooth to a sufficiently high precision.

\textls[-15]{Another important implication is related to the entropic state of the Universe. This aspect is crucial for understanding the naturality of the initial conditions and evolution of the Universe.
A puzzling property of the observed Universe is that, whereas the energetic degrees of freedom (photons, baryons, electrons, and neutrons) were at thermal equilibrium in the very early Universe, i.e., at maximum entropy, gravity appears to be at very low entropy, as evidenced by the shallow potential wells, 
$\zeta_{rms}=\mathcal{O}(10^{-5})$, at least at the recombination era. Such spatial homogeneity level is atypical for systems governed by long-range attractive forces like gravity. The penultimate highest entropy state for such a system --- when other non-gravitational interactions can be ignored --- is matter clumped in black holes (with typical $\zeta_{rms}\lesssim 1$). Ultimately, even these black holes are evaporated after stupendously long time scales into an entropically favorable state --- diffused thermal~radiation.}

Since the fundamental interactions of Nature are time-reversal-symmetric (except for a small violation of this symmetry in the weak interaction sector that is too weak to account for the observed Arrow of Time) it then follows that the observed global 
time-asymmetry in Nature is 'explained' by very special initial conditions of our Universe --- the Past Hypothesis, e.g.,~\cite{Wallace2011,Carroll2014}. Rather ironically, the Past Hypothesis simply posits that our Universe started at a very low entropy state compared to a 'typical' Universe, only to explain its current statistically unlikely state by the existence of previous --- even more unlikely --- states. Standard estimations, e.g.,~\cite{Dyson2002,AlbrechtSorbo2004}, imply that the entropy in the pre-inflationary phase of our Universe did not exceed $10^{10}$ by much, and essentially the entire $S\sim 10^{88}$ in the form of cosmic microwave background (CMB) photons and background neutrinos is conjectured to have formed via aggressive and brief dissipative `reheating' phase right at the end of the inflationary expansion era; this scenario places the initial state of our Universe in a very special and unique spot in the space of all possible initial configurations, given the current state of the Universe and the time-symmetry of the fundamental interactions.

Central to these considerations is the realization that the observable Universe appears to be a closed system to a very good approximation, due to the shallowness of gravitational wells at the horizon, which hinder significant inflow or outflow of entropy, i.e., material (rather than gravitational) degrees of freedom. Since the observable Universe can be approximated as a closed system to a very good precision it then follows by virtue of the Second Law of Thermodynamics that entropy cannot decrease over time. Consequently, the emerging picture is that for some reason the low entropy in the gravitational degrees of freedom is also the reason that our Universe can be effectively considered a closed system with a non-decreasing entropy. 
{As mentioned above, this perspective is closely related to the Past Hypothesis and to modern discussions of the thermodynamic Arrow of Time in cosmology, which attribute the observed time asymmetry to special low-entropy initial conditions rather than to time-asymmetric microscopic laws~\cite{Wallace2011,Carroll2014}.}
According to the Small Patch Hypothesis, this low-entropy reflects our limited observable volume; in the working picture adopted here, on scales far beyond the Hubble radius the typical amplitude of metric perturbations is of order unity, so that spacetime is highly inhomogeneous and (by analogy with black-hole dominated configurations) plausibly close to a maximum gravitational-entropy state. 
Related arguments concerning the entropic interpretation of cosmological initial conditions and the role of horizons in defining effective closed systems have been discussed in, e.g.,~\cite{Dyson2002,AlbrechtSorbo2004}.
This identification between strong large-scale inhomogeneity and high gravitational entropy is heuristic, since no universally accepted microscopic definition of gravitational entropy exists away from horizons.

The paper is organized as follows. 
In Section~\ref{sec2} a list of challenges and open questions 
faced by the inflationary scenario are summarized.
In Section~\ref{sec3} a non-inflationary approach to the flatness and horizon problems is discussed in the context of a minuscule observable Universe embedded in a much larger and conformally older spacetime. 
In Section~\ref{sec4} the proposed power spectrum --- a single realization of a class of infinitely many power spectra --- is laid out along with its implications. In Section~\ref{sec5} the phase coherence argument, 
and the gaussian and adiabatic perturbations are discussed in the context of the Small Patch Hypothesis. This work is summarized in Section~\ref{sec6}.
The proposed logarithmically running spectrum is confronted with cosmological observations in Appendix~\ref{appA}.

\section{Conceptual Challenges and Open Questions in Inflation}\label{sec2}

Inflation was originally introduced to provide a dynamical explanation of several puzzling features of the Hot Big Bang framework, including the global spatial flatness, smoothness over scales that could not have been in causal contact assuming sub-luminal expansion and the absence of relic magnetic monopole remnants from phase transitions in the early Universe, e.g.,~\cite{Guth1981,Linde1982,Hawking1982,MukhanovChibisov1981,
Starobinsky1982}. It was soon after realized that inflation explains the observed near scale-invariant, adiabatic, and approximately Gaussian statistics of primordial perturbations as well. In addition, evolving quantum vacuum fluctuations to macroscopic perturbations stretched to astrophysical scales, inflation offered a compelling underlying microphysical mechanism behind the observed large-scale structure. Clearly, inflation has rightfully become the dominant early-Universe paradigm. 

Nevertheless, alongside these attractive features, a number of stability and naturalness issues, and open conceptual difficulties still linger. Importantly, these difficulties do not apply uniformly to all inflationary realizations; rather, different challenges are associated with different classes of models, and no single known class avoids all of them simultaneously without introducing additional assumptions. 

First, inflation generically predicts an eternally self-reproducing spacetime in which the inflating volume diverges, implying that probability estimates require regularization. The latter is effected via an arbitrary choice of cutoff or measure, and because there is no unique prescription, different measures --- such as proper-time cutoffs, scale-factor cutoffs, causal-patch measures, or light-cone measures --- lead to widely varying predictions for cosmological observables. This is the well-known ``measure'', ``youngness'' and volume-weighting problems, e.g.,~\cite{Guth2007,Freivogel2006,Bousso2006}. These issues are most acute in models that admit eternal inflation (e.g., false-vacuum/landscape and some large-field models).

Second, successful inflation requires the inflaton potential to be extremely flat. Such flatness is unstable in the presence of radiative corrections, which generically drive the slow-roll parameters away from the very small values required for successful inflation, giving rise to the ``$\eta$-problem'' and severe fine-tuning, unless specific symmetries are imposed~\cite{Stewart1997,Baumann2009}, e.g., ``shift symmetry''. Imposing such symmetry constraints solely to stabilize the slow-roll regime can by itself be viewed as fine-tuning. Supergravity and string-inspired realizations of inflation often face even more severe obstacles, since moduli field stabilization may compromise the required flatness of the potential. Moreover, inflation generically requires finely tuned initial conditions; the inflaton must begin sufficiently homogeneous over super-Hubble regions and must start sufficiently high on its potential for successful slow roll, both conditions appear non-generic and fine-tuned~\cite{GoldwirthPiran1992,GibbonsTurok2008}. These issues have been argued to be particularly pronounced in small-field models which can require special initial conditions to enter and sustain the slow-roll regime.

A related difficulty is the ``trans-Planckian problem''. The large-scale perturbations observed in the CMB originate from wavelengths which, when traced back to the early inflationary stage, correspond to super-Planckian energy scales. Their initial conditions may therefore be sensitive to unknown ultraviolet physics~\cite{BrandenbergerMartin2001,BrandenbergerMartin2001b,NiemeyerParentani2001,BrandenbergerMartin2005}. 
This issue is largely model-independent and affects essentially all inflationary scenarios that invoke a sufficiently long period of accelerated expansion, insofar as observable modes can originate from sub-Planckian physical wavelengths when evolved backward in time.
In addition, CMB observations, e.g.,~\cite{PlanckInflation2018}, have already ruled out a large class of simple inflationary models, narrowing the viable parameter space and raising the question of whether the remaining successful scenarios reflect deep physical principles or merely residual model flexibility.

Thus, inflation --- proposed as a resolution to classical fine-tuning puzzles --- appears to require its own tunings. In addition, certain realizations of the inflationary scenario predict eternal inflation, leading to a multiverse with infinitely many causally disconnected spacetime regions between which fundamental constants vary~\cite{Guth2007,Bousso2000}. This exacerbates the measure problem and degrades the predictive power of inflation. Finally, the Borde-Guth-Vilenkin theorem demonstrates that even eternally inflating spacetimes are geodesically past-incomplete. Therefore, Inflation cannot be the fundamental beginning and additional pre-inflationary physics is required~\cite{BGV2003}. This result applies broadly to all inflationary spacetimes satisfying an average expansion condition, independently of the specific inflaton~potential.

Overall, these issues suggest that whereas inflation successfully explains observed features of the Universe, it is theoretically and conceptually incomplete~\cite{Ijjas2014}. Alternative frameworks are therefore worth exploring --- not necessarily to replace inflation, but to clarify which observed features are uniquely inflationary predictions and which may arise more generically. In particular, inflation provides a 
dynamical route to homogeneity, flatness, Gaussianity, and adiabaticity, whereas the scenario proposed in the present work explores whether some or all of these observational imprints may instead follow from geometric reasons, i.e., the observable Universe is a tiny fractional volume of a much larger {and conformally older} spacetime. This idea is referred to here as the Small Patch Hypothesis, i.e., the proposal that several features commonly attributed to inflationary dynamics may instead reflect the possibility that our observable cosmological domain represents only a tiny fraction of a much larger causally connected spacetime. Indeed, some analyses argue that as cosmological data improve, support for inflationary models strengthens~\cite{Chowdhury2019}, whereas others suggest otherwise. The present work is motivated by this latter possibility.

\section{The Flatness and Horizon Problems in a Very Large Universe}\label{sec3}

In the standard cosmological model, the 'horizon problem'~\cite{Rindler1956,Misner1969} arises because, in a Universe with a finite conformal past, sufficiently remote regions of the CMB sky could never have been in causal contact prior to recombination given the limited time available for signals to propagate since the Big Bang and assuming sub-luminal expansion, yet they exhibit nearly identical temperatures, to a one part in $10^{5}$ precision level. Similarly, the 'flatness problem'~\cite{Dicke1970} refers to the (seemingly) extreme fine-tuning required for the spatial curvature parameter $\Omega_{k}$ at, e.g., the grand unification theory (GUT) epoch, 
to not end up being of order unity today, e.g.,~\cite{Weinberg2008,LiddleLyth2000}. Both are classical problems of the Hot Big Bang~model.

The inflationary paradigm offers a dynamical resolution of both puzzles at once. Specifically, a brief phase of exponential expansion causes initially small, causally connected regions to be stretched far beyond the present Hubble radius. In addition, any pre-inflationary nonvanishing spatial curvature scale is stretched to super-Hubble scales rendering the observable patch of post-inflationary space essentially flat, with 
$|\Omega_{k}|\ll 1$ effectively.
As a result, the observable Universe is both causally connected and indistinguishable from spatially flat, e.g.,~\cite{Guth1981,Linde1982}.
Significantly, inflating the spatial curvature beyond the horizon comes at a cost; all pre-inflationary matter in the observable Universe is expelled beyond the horizon and new post-inflationary matter has to form. Inflation achieves this feat by postulating the decay of the inflaton field shortly after inflation via an extremely dissipative highly model-dependent 'reheating' phase, that results in the generation of $\mathcal{O}(10^{88})$ degrees of freedom in a very short time period, further exacerbating the fine-tuned initial conditions required for the start of inflation; the latter must have started at a very unique --- low entropy --- state.

In contrast, in a conformally past-complete or past-eternal cosmological scenario, or at least if the conformal scale of the entire hosting Universe is much larger than that of the observable Universe, the usual sharp formulations of the horizon and flatness problems are softened; causal contact and a large curvature radius can arise simply because the parent spacetime has a much longer conformal age and much larger characteristic size than the observable domain. In that sense the ``problems'' are shifted from the observable patch to questions about the origin and properties of the parent Universe, which we do not attempt to address here. In fact, these are largely irrelevant to the present discussion; the only properties of the hosting Universe that really matter is that its causally connected domain is much larger than that of the observable Universe. Such a much older Universe may plausibly be closer to equilibrium on sufficiently large scales, or at least lack large entropy gradients relevant for the observable patch. Importantly, this is not much different than the situation portrayed by the inflationary scenario where the spatial curvature radius and causally-connected scale are much larger than the observable~Universe.

Crucially, inflation also provides a quantum-mechanical origin, e.g.,~\cite{MukhanovChibisov1981, Sasaki1983, Mukhanov1992,PolarskiStarobinsky1996,KieferPolarski1998,LiddleLyth2000, Durrer2008,CampoParentani2008,MartinVenin2016}, for the observed slightly red-tilted spectrum of metric perturbations of the scalar type. The nearly flat spectrum was contemplated on general physical grounds a decade earlier~\cite{Harrison1970, PeeblesYu1970, Zeldovich1972}, so in this sense inflation actually retrodicted this spectrum, and --- admittedly --- predicted its gaussianity and adiabaticity.

If spacetime is conformally past/future-eternal, or at least very large, while the observable Universe is only a finite, causally-connected region embedded within a much larger hosting Universe, then in such a scenario, the curvature radius $l_{\text{curv}}$ is typically much larger than the extent of the observable Universe. The latter is set by the cosmological constant, $R_{\Lambda}\sim \Lambda^{-1/2}$. Consequently, the observed spatial flatness and causal-connectedness are expected geometric consequences of residing within a tiny fraction of an enormously large hosting Universe.

At this point it is important to emphasize that the Small Patch Hypothesis is fundamentally a statement about observational boundedness rather than about the detailed global properties of the hosting spacetime. If dark energy is indeed a true cosmological constant, $\Lambda>0$, then the observable Universe is eternally bounded from above by the asymptotic de~Sitter horizon scale $\sim \Lambda^{-1/2}$, independently of the cosmic time at which observations are made. In particular, whether observations are carried out $10$~Gyr, $10^{4}$~Gyr, or $10^{50}$~Gyr after the Big Bang, the maximum comoving scale that can ever be accessed remains finite and fixed by $\Lambda$. This bound is purely geometric, and is entirely independent of inflationary dynamics, anthropic considerations, or assumptions regarding the homogeneity or isotropy of the spacetime on scales larger than the observable~domain.

In this sense, the horizon and flatness problems do not represent fine‐tuning puzzles but rather reflect observational selection effects; we only observe a small, causally connected region of a vastly larger spacetime. This geometric interpretation is fully consistent with 
the Small Patch Hypothesis put forward in the present work.

As a consequence, even if the hosting Universe is highly inhomogeneous on sufficiently large scales, there always exists a regime in which the observable patch — bounded by the de~Sitter horizon — appears smooth, homogeneous, and well described by linear perturbation theory, provided $\Lambda$ is sufficiently large. The Small Patch Hypothesis therefore does not require the hosting Universe itself to be homogeneous, isotropic, or dynamically smoothed; rather, it exploits the fact that positive $\Lambda$ enforces an unavoidable observational cutoff that regulates which Fourier modes of the primordial spectrum can ever be probed. In this sense, the focus of the present framework is the observable Universe and its causal boundaries, not the detailed structure of the spacetime beyond them, much like the inflationary paradigm does not provide any concrete predictions about the spatial curvature and causal scales of the much larger spacetime in which the observable Universe is embedded.

Thus, both the horizon and flatness problems can be addressed not only through dynamical inflation, but also as geometric and causal features of a tiny observable Universe embedded within a very large (and possibly past-eternal) spacetime. In this framework, the inflationary expansion phase is unnecessary for explaining causal-connectedness and near-flatness, though additional mechanisms or explanations would be needed to account for the origin of primordial perturbations.

It is worth reiterating and emphasizing the key conceptual distinction between the present framework and the inflationary paradigm. In inflation, a specific high-energy phase of accelerated expansion pushes modes outside the horizon at very early times, relying on unknown microphysics allegedly operating far above experimentally tested energy scales. In contrast, the Small Patch Hypothesis explains the exclusivity of observable scales geometrically; a positive cosmological constant enforces a permanent, epoch-independent causal bound on what any observer can ever access, irrespective of the detailed early-time dynamics. Whereas inflation renders certain modes unobservable by expelling them from the horizon, the present framework posits that such modes are never observable in the first place. This distinction shifts the explanatory burden from speculative ultraviolet physics to a low-energy parameter, $\Lambda$, that is directly measured in the late Universe, and reframes large-scale smoothness as a consequence of causal accessibility rather than of early-Universe~dynamics.

In addition, inflation is driven by a high-energy vacuum energy allegedly 
contributed by the inflaton field. The latter is foreign to the standard model of 
particle physics, and a Universe eternally dominated by vacuum energy is inconsistent with the Universe we observe. As mentioned above, all this implies that the current entropy associated with 
material degrees of freedom, $S\sim 10^{88}$, must have formed by the decay products of the inflaton field, implying that the Universal entropy at the inflationary epoch was $S_{inf}\sim 10^{10}$. 
This by itself represents a stupendously huge fine-tuning of the pre-inflationary initial conditions. 
Remarkably, this exceptionally severe fine-tuning simply does not exist within the Small Patch scenario.

\section{A Logarithmic Primordial Spectrum and the Small Patch Hypothesis}\label{sec4}

Within the framework of inflation, the primordial power spectrum of scalar-type metric perturbations is 
\begin{eqnarray}
P_{s}(k)\propto\frac{H_{{\rm inf}}^{2}}{\epsilon m_{pl}^{2}}\propto \frac{\rho_{{\rm inf}}}{\epsilon\rho_{pl}}, 
\end{eqnarray}
where $H_{{\rm inf}}$ is the (nearly constant) expansion rate during the inflationary era, $m_{pl} \propto G^{-1/2}$ is the Planck mass, 
$\epsilon$ is a model-dependent 'slow-roll' parameter (gauging the flatness of the inflaton potential), $\rho_{pl}\sim m_{pl}^{4}$ and $\rho_{{\rm inf}}$ are the Planck energy density and the (vacuum-like) energy density at inflation, respectively, and in the last step in Equation~(1) the Friedmann equation has been employed. 
Here, scalar metric perturbation should be understood concretely as the gauge-invariant Bardeen potential $\Phi$ (or, equivalently, the comoving curvature perturbation $\zeta$ in the super-horizon limit), e.g.,~\cite{LiddleLyth2000, Durrer2008,Weinberg2008}. 

Although the default assumption in the standard cosmological model framework is that $P_{s}(k)=A_{s}(k/k_{0})^{n_{s}-1}$, the implications of a wide range of alternative spectral dependencies of $P_{s}(k)$ have been explored in the literature, e.g.,~\cite{Adams2001,Flauger2010,Meerburg2014,Achucarro2014,MirandaHu2014,Albrecht2014,HuTorrado2015,Fergusson2015,Palma2015,Fergusson2015b,Cai2015,Mooij2015,Chen2016,Chen2016b,Zeng2019,Beutler2019,Yoshiura2020,Domenech2020,Yoshiura2020b,Sohn2024,Yang2023,Naik2025}.

Motivated by the Small Patch Hypothesis {(and for illustrative purposes only)}, we postulate a certain power spectrum that nearly overlaps with the standard red-tilted power spectrum over intermediate scales but markedly deviates on super-horizon and very small scales and ultimately diverges on scales, $k_{c}^{-1}$, many orders of magnitude larger than the Hubble scale, 
$H_{0}^{-1}$. Accordingly, the running of $P(k)$ that captures the expected behavior on observable scales is taken as a phenomenological parameterization in the same sense that the Harrison-Zeldovich (HZ) spectrum was initially proposed as a phenomenological ansatz,~\cite{Harrison1970, PeeblesYu1970, Zeldovich1972}. It is representative of infinitely many other possible power spectra that approximate the red-tilted spectrum on observationally accessible scales and diverge on ultra-large scales, thereby determining the size of the hosting Universe. \textls[-15]{{All the consequences of the Small Patch Hypothesis readily follow in case that the hosting Universe has infinite causal size. 
The purpose of the present section is to illustrate their validity even in the case that the observable Universe is embedded within a very large but finite hosting Universe. The size of the latter is plausibly determined by $k_{c}^{-1}$ in which its gauge-invariant description becomes singular and consequently breaks down. Therefore, neither the divergence of $P(k)$ at a finite $k_{c}$ in general, nor the particular logarithmically running $P(k)$ discussed in the present section are  
essential to the plausibility of the Small Patch Hypothesis and the integrity of the corresponding arguments, but are rather primarily meant as a demonstration of its applicability under more constraining~conditions.}} 

Since infrared behavior of the Newtonian potential $\Phi(k)$ is gauge-dependent on super-horizon scales, e.g.,~\cite{Bardeen1980,Bruni1997,LiddleLyth2000, Weinberg2008, Durrer2008}, and because cosmological inference is performed using the conserved curvature perturbation $\zeta(k)$ anyway, we consider the logarithmic running as a property of the gauge-invariant primordial curvature spectrum rather than of the Newtonian one. The phenomenological ansatz is therefore
\begin{equation}
P_{\zeta}(k)=\frac{A_{s}}{1-(n_{s}-1)\ln(k/k_{0})}.
\end{equation}
For a single perfect fluid with constant equation of state $w=p/\rho$, where $\rho$ and $p$ are the energy density and pressure, and $k\ll aH$, $\Phi(k)$ and $\zeta(k)$ are related 
via $\Phi(k)=\frac{3(1+w)}{5+3w}\zeta(k)$.
For realistic multi-component evolution with time-dependent $w(\eta)$ [where $\eta$ is the conformal time, which is the time coordinate capturing the causal --- rather than cosmic --- reach of a given spacetime], this relation generalizes to a spacetime-dependent transfer function,
$\Phi(k,\eta)=T_{\Phi}(k,\eta)\zeta(k)$,
where $T_{\Phi}(k,\eta)=\frac{3}{5}$ on large scales during matter domination (MD).  
This ensures that the logarithmic form of $P_{\zeta}(k)$ propagates forward through standard cosmological evolution in exactly the same way as the conventional red-tilted power-law spectrum on all observable sub-horizon scales. In the following $P_{s}(k)$ actually refers to 
$P_{\zeta}(k)$.

Since observationally $n_{s}-1\approx -0.03$, the power spectrum diverges at sufficiently low $k$ values. Specifically, this divergence formally takes place at a `critical' value,
\begin{eqnarray}
l_{c}=l_{0}e^{1/(1-n_{s})}. 
\end{eqnarray}
At around the pivot scale, $l_{0}=20$ Mpc, we infer $n_{s}=0.968$, and so $l_{c}=\mathcal{O}(10^{39})$ cm, at which scale gravitation wins out all other interactions and super-horizon metric perturbations become $\mathcal{O}(1)$ fluctuations.  
In a $\Lambda$CDM Universe asymptoting to $l_{\Lambda}=\left(\frac{\Lambda}{3}\right)^{-\frac{1}{2}}$~\cite{KraussScherrer2007}, requiring that perturbations remain linear across the observable domain yields
\begin{eqnarray}
1-n_{s}\ll\frac{1}{\ln(l_{\Lambda}/l_{0})}. 
\end{eqnarray}
Using the observationally inferred $l_{\Lambda}\sim\mathcal{O}(10)$ Gpc, 
and $l_{0}=0.02$ Gpc gives
\begin{eqnarray}
0.84\ll n_{s}<1, 
\end{eqnarray}
consistent with the measured value $n_{s}= 0.968$. Thus, red-tiltedness emerges here as a geometric consistency condition rather than a slow-roll prediction {(concavity of the inflaton potential)}, conditioned on the logarithmic ansatz, Equation~(2), being valid over at least %\sout{ten} 
{a dozen} decades in $k$ beyond the observed window. 

Requiring that $P_{s}(k)\ll 1$ across observable scales yields the condition
\begin{eqnarray}
\exp\left(\frac{1-A_{s}}{1-n_{s}}\right)\gg\frac{l_{\Lambda}}{l_{0}}. 
\end{eqnarray}
For $l_{\Lambda}/l_{0}>1$, this implies either $(A_{s}<1,n_{s}<1)$ or $(A_{s}>1,n_{s}>1)$. In the observationally relevant regime where $A_{s}\ll 1$, one necessarily obtains $n_{s}<1$. Anthropically motivated considerations such as~\cite{TegmarkRees1998} reinforce this requirement within the same class of spectra that grow toward the infrared.

Whereas Equation~(2) departs from a simple power law sufficiently far 
from $k_{0}$, the limited range of scales probed by the CMB and clustering data does not provide sufficient leverage to significantly detect a departure. A preliminary Monte Carlo Markov Chains (MCMC) comparison using current CMB and large-scale structure data confirming this behavior appears in Appendix~\ref{appA}. At present, the logarithmic model should therefore be regarded as a concrete representative of a broader family of infrared-enhanced spectra consistent with existing data.

As pointed out above, the power spectrum described by Equation~(2) is just one illustrative example. Alternative spectra with IR divergence at a finite $k_{c}$ that approximate the red-tilted power law over scales at the vicinity of $k_{0}$ are abundant.
For example, a straightforward generalization of Equation~(2) is 
$\tilde{P}(k)=A_s\Big[1-\frac{n_s-1}{p}\ln(k/k_0)\Big]^{-p}$ (with $p>0$) which is approximated as a red-tilted power law in the neighborhood of $k_{0}$, i.e., $\tilde{P}(k)\approx A_s(k/k_0)^{n_s-1}$, but has an infrared divergence at $k_{c}=k_{0}\exp(\frac{p}{n_{s}-1})$. For a fixed observationally inferred $n_{s}$ increasing $p$ increases $k_{c}$. For the divergence scale 
$l_{c}\sim l_{0}e^{\frac{p}{1-n_{s}}}$ to exceed the asymptotic de Sitter scale we must have $p\gtrsim 0.2$. For reference, Equation~(2) corresponds to the case $p=1$.

The role of the logarithmic (or finite-$k_c$) infrared divergence is purely to demonstrate existence; for any primordial spectrum that becomes highly nonlinear at a finite scale $k_c>0$, a sufficiently large positive $\Lambda$ can be found such that the observational cutoff 
$l_{\Lambda}\sim\Lambda^{-1/2}$ lies entirely within the linear regime, ensuring that the observable Universe appears smooth and homogeneous over the observable range of scales (regardless of inflation) even if the hosting spacetime becomes highly inhomogeneous on larger scales.

\section{Statistical Properties of Primordial Perturbations in the Small Patch~Hypothesis}\label{sec5}

\textls[-15]{The structure in the observed CMB anisotropy and polarization is consistent with phase-coherent, Gaussian, and adiabatic primordial perturbations. In inflation, these properties arise dynamically; quantum fluctuations of the inflaton and metric fields are stretched to super-Hubble scales where they ``freeze'' and become classical realizations of a nearly Gaussian stochastic field~\cite{Mukhanov1992,PolarskiStarobinsky1996,KieferPolarski1998,CampoParentani2008,MartinVenin2016}. Once such a mode later re-enters the horizon ($k=aH$) it resumes plasma oscillations with a well-defined phase shared across all realizations of that wavenumber, thereby generating the observed acoustic peak structure in the CMB~\cite{Albrecht1996,Magueijo1996,Turok1996,Dodelson2003}. In this sense, inflation provides a well-structured route to classicality and coherent initial conditions.}

In the proposed (possibly classical) non-inflationary scenario, the parent Universe is assumed to be much larger and conformally older (perhaps even past-eternal) than the asymptotic de Sitter scale of the observable Universe, and scalar-type metric perturbations are assumed to have existed on all wavelengths from arbitrarily small to extremely large scales prior to the Big Bang in the observable Universe. This hosting Universe is assumed to be in equilibrium or near-equilibrium. In fact, this is expected rather than merely assumed provided that this parent Universe is much older than the observable Universe. In such a Universe, perturbations remain effectively static as long as their physical wavelengths exceed the Hubble radius.  
As the Hubble horizon grows, progressively smaller $k$-modes 
satisfy the condition $k=aH(\eta)$ and begin their dynamical evolution 
upon in-bound crossing the horizon.  
Thus, each comoving scale begins oscillating at its horizon-entry time, 
and all realizations of that scale do so synchronously.  
In this sense --- unlike in the inflationary scenario --- phase coherence emerges not from freeze-out and re-entry, but from a geometric, entry-only that results  inherently from the fact that the observable Universe occupies a tiny volume within a much larger and conformally older spacetime.  
Here, a conceptually different mechanism is at work; whereas inflation derives coherence from quantum field dynamics in a nearly de Sitter space, the Small Patch Hypothesis attributes coherence to geometric boundaries of the observable Universe at any given time, with an implicit assumption that the pre-existing perturbations are dominated by the same growing adiabatic mode that is selected dynamically in simplest inflationary models.

Gaussianity is an important and distinguishing empirical property of the primordial perturbations, which is normally considered as yet another hallmark of the simplest models of inflation.  
Historically, long before inflation provided a dynamical mechanism for the generation of Gaussian perturbations, a few independent arguments pointed to the Gaussian random field as the natural statistical description of primordial fluctuations, based on random phases, the central limit theorem, and maximum entropy reasoning~\cite{Silk1968,Zeldovich1970,Doroshkevich1970,Peebles1973,Peebles1980,Bardeen1980,BBKS1986}.  
Inflation later provided a structured microphysical framework from which this statistical prior naturally emerges; in the simplest slow-roll single-field models, primordial non-Gaussianity is strongly suppressed and in fact depends on inflationary slow-roll parameters~\cite{Maldacena2003,Creminelli2004,Weinberg2008,Komatsu2010}.  
Current observational constraints are consistent with this expectation~\cite{Planck2019NG}.   
Likewise, the same class of inflationary models predicts purely adiabatic 
initial conditions, consistent with the strong observational limits 
on primordial isocurvature modes~\cite{Gordon2001,PlanckInflation2018}.

More general inflationary scenarios (e.g., multifield, non-canonical, 
or features in the potential) can produce detectable non-Gaussianity and/or observable isocurvature 
perturbation modes~\cite{Lyth2003,Bartolo2004,Langlois2008,Chen2010,Chen2010b,Byrnes2010,Meerburg2019}.

In contrast, in the proposed Small Patch Hypothesis, Gaussianity and adiabaticity arise from a different physical principle; we postulate (and in fact naturally expect following the argument outlined above) that all scales above the observable horizon are effectively static and can be modeled as a coarse-grained equilibrium state that maximizes entropy subject to a restricted number of macroscopic constraints.  
Conditionalizing only on fixed mean and two-point function of the perturbation field and imposes no additional constraints, the maximum-entropy probability distribution is uniquely Gaussian~\cite{Jaynes1957} (much like the Maxwell-Boltzmann distribution in kinetic theory). In this sense, and within these assumptions, Gaussianity appears as a maximum-ignorance prior for the external field configuration, rather than as a consequence of a specific high-energy microphysics and fine-tuned suppression of the self interaction of an inflaton field.

The same reasoning applies to adiabaticity. 
If, in addition, one assumes that the macroscopic constraints involve only the total energy density of each species, then in a maximum-entropy configuration the relative number densities of radiation, baryons, cold dark matter, neutrinos, etc., are spatially uniform and entropy perturbations vanish. Within this restricted thermodynamic picture, the absence of primordial isocurvature modes is then interpreted as a property of the equilibrium parent Universe. 
In practice, admittedly, we impose adiabaticity as part of the definition of the external equilibrium configuration, rather than rigorously deriving it from a detailed microscopic treatment of gravitational entropy, or rather gravitational entropy maximization principle.

In summary, phase coherence, Gaussianity, and adiabaticity --- often viewed as fingerprints of inflation --- can instead be interpreted broadly as emergent consequences of observing only a limited causal region of an otherwise much larger (and conformally older) Universe in equilibrium. Whereas inflation derives these properties from early-Universe field dynamics (in a model-dependent fashion), the Small Patch Hypothesis attributes them squarely to geometric horizon-entry and the thermodynamic properties of a frozen, maximum-entropy exterior, at the cost of elevating specific assumptions about the parent equilibrium state to the status of {plausible} macroscopic constraints. 

\section{Summary}\label{sec6}

The phenomenal success of the inflationary scenario can largely be attributed to the fact that according to inflation the observable Universe occupies a tiny volume of a much larger causally connected Universe. In this setup the infamous horizon and flatness problems trivially go away. Inflation achieves this by explaining how the observable Universe could have undergone a brief $\sim$60--70 e-fold of super-luminal exponential expansion.
This idea is conventionally realized by postulating the existence of an inflaton field (or fields) dominated by its potential energy, thereby mimicking a vacuum-like energy density that implies an exponential expansion. Quantum perturbations of the inflaton generically induce gaussian adiabatic matter perturbations that are described by a slightly red-tilted power spectrum. 

However, any alternative scenario which relies on the idea that the observable Universe is a tiny fraction of a much larger causally connected parent Universe automatically addresses the classical horizon and flatness problems. The latter problem goes away because typically (at least in the case of a globally closed Universe) the curvature radius defines the size of the Universe. If the latter is much larger than the observable Universe then the latter is practically indistinguishable from a spatially flat one. We therefore adopt a Small Patch Hypothesis throughout this work; the observable Universe is embedded in a much larger causally connected Universe which is sufficiently (conformally) older than the observable Universe. This latter property explains the apparent horizon problem away. 

Consistent with this hypothesis (but not required by it) we adopt an 
ansatz power spectrum that logarithmically runs with scale over cosmological and ultra-Hubble scales until it diverges in the far infrared regime, on scales a trillion times larger than the Hubble scale, or larger. This divergence essentially delineates the size of the hosting Universe {because on larger scales the gauge-invariant description of the perturbed hosting Universe breaks down}. On cosmologically observable scales the proposed power spectrum is well-approximated by the standard slightly red-tilted power law spectrum. The particular power spectrum we adopted in the present work could be replaced by any other spectrum insofar it has a support on a few decades of scales beyond the Hubble scale and provided that it can be well-approximated by a red-tilted power law on the observationally available window of scales; the specific lower bound on $\Lambda$ and the inferred divergence scale are therefore conditioned by this chosen functional form, but similar reasoning would apply to other spectra that grow toward the infrared, including the slightly red-tilted generic inflationary power law. In the latter case the power spectrum formally diverges on infinitely large scales and it is easier to achieve homogeneity of the observable Universe. In the present work we considered 
a power spectrum that diverges on (sufficiently) large and finite scale in order to demonstrate the validity of the Small Patch Hypothesis even under more restrictive~conditions.

The CMB only samples 3--4 decades in scale (multipoles $2<\ell<10^{4}$), providing insufficient leverage to rule out the proposed logarithmically running $P_{\zeta}(k)$ given its observationally inferred slow spectral slope. There is simply not enough information in current data to significantly distinguish an inflationary-motivated slightly red-tilted power law from the non-inflationary logarithmic spectrum considered here. It is assumed that this spectrum can be extrapolated to super-horizon scales before diverging on scales trillions of times larger than the observable Universe, where the very notion of linear perturbations around a smooth background breaks down.

The cosmological implications of such a spectrum are far-reaching in the context of the classical Hot Big Bang horizon and flatness problems. In particular, the inferred size of the cosmological constant $\Lambda$ sets the asymptotic de Sitter horizon scale of the observable Universe, which in this picture is simply too small to access the large-amplitude regime of scalar metric perturbations. 
Crucially, this bound is independent of the detailed structure, age, or degree of inhomogeneity of the hosting Universe; for sufficiently large $\Lambda$, any observer --- regardless of cosmic 
epoch --- necessarily inhabits an observationally finite, causally bounded region that can appear smooth and homogeneous even if the parent spacetime is highly inhomogeneous on much larger scales.
This restricted observational window renders the observable Universe a ``zoom-in'' patch on a much larger spacetime, reminiscent of the role played by primordial inflation. However, whereas inflation provides a dynamical smoothing mechanism that relies on unknown high energy physics (perhaps a trillion times higher than our experimentally established standard model of particle physics), the present framework is fundamentally geometric; homogeneity and flatness arise because we observe only a tiny fraction of a far larger Universe. Unlike inflation, this does not invoke a new high-energy scalar sector with a finely tuned potential, but instead assumes the existence of a much larger parent spacetime whose global properties (including its entropy content) are not explained here --- the proposed hypothesis is largely independent of any assumptions about the parent Universe, other than 
the latter is much larger and conformally older than the observable Universe. 
In this sense, the Small Patch Hypothesis is not a phenomenological substitute for inflation but a conceptual alternative explanation of the same empirical facts --- flatness, causal connectedness, (and provides a tentative explanation for Gaussianity, adiabaticity, and near scale invariance) --- showing that they can arise from causal boundedness and entropy considerations alone, without invoking early-Universe dynamics or speculative microphysical sectors.

Crucially, as with the Harrison-Zeldovich-Peebles spectrum that was historically proposed based on phenomenology alone (before it was shown to naturally emerge from inflation), the logarithmic running considered here is employed phenomenologically, not derived from a microscopic field theoretic model. It should be 
emphasized that any power spectrum that is approximated by the observed red-tilted power law in the observable Universe is consistent with the Small Patch Hypothesis. It may have infrared divergence on finite super-Hubble or infinitely large scales without having any observable signature.
In the present proposal the observed gaussianity and adiabaticity of perturbations are motivated as follows. If the perturbation field outside our Hubble volume is effectively frozen (a natural assumption in a conformally old or eternal host Universe), we model it as (or arbitrarily close to) a coarse-grained maximum-entropy configuration. Assuming that only a fixed mean and rms fluctuations are conditioned, Gaussianity follows as the maximum-entropy (``maximum ignorance'') distribution, while adiabaticity corresponds to the absence of spatially varying relative abundance of the various species. In this sense, Gaussian and adiabatic initial conditions may be understood as capturing our ignorance about the external state --- indeed a maximum ignorance state --- rather than as signatures of a particular high-energy inflaton sector. Admittedly, this argument remains somewhat heuristic, because it does not rely on a fully specified microscopic definition of gravitational entropy in general spacetimes. On the other hand, and this should be appreciated as well, global macroscopic properties of the Universe could well be emergent with no fundamentally underlying microscopic mechanism.

A well-known challenge in cosmology is that large-scale homogeneity corresponds to a very low gravitational entropy state, S. While matter (dominated by relativistic relics) contributes $S\sim 10^{88}$, this remains vastly smaller than the de Sitter entropy $S_{dS}\sim 10^{122}$. Thus, soon after the Big Bang --- and still today --- the gravitational entropy of the Universe is anomalously 
low {even when the huge estimated entropy of supermassive black holes in centers of galaxies is accounted for}. In standard inflationary cosmology this is sometimes attributed to finely tuned initial conditions (the Past Hypothesis), requiring an exceptionally low pre-inflationary entropy followed by a highly dissipative reheating phase. In contrast, the present work suggests that the observed low gravitational entropy reflects the fact that we only probe scales where spatial metric inhomogeneities remain small. On much larger scales --- beyond $\sim$$\Lambda^{-1/2}$, and even more so near the divergence scale implied by logarithmic running (or any other monotonic extrapolation of the observed power law to ultra-Hubble scales for that matter) --- space is expected to become highly inhomogeneous and the gravitational entropy plausibly approaches a value much closer to the maximum allowed by the global constraints, by analogy with black-hole dominated configurations. From this viewpoint, the cosmological Arrow of Time can be emergent and local, not fundamental; it characterizes the low-entropy pocket we inhabit, 
not the global state of the vastly larger Universe, although a fully quantitative account of this connection would require a more precise and widely accepted notion of gravitational entropy. However, such a universally accepted definition of gravitational entropy indeed does not exist at the present time.

Current observations lack the dynamic range needed to distinguish between the proposed logarithmic spectrum (or any other spectrum that is sufficiently accurately mimicked within the observable Universe by the observed red-tilted power law) and the standard inflationary power law, but this stalemate situation could end when new data comes. In particular, 
21-cm cosmology of the Dark Ages ($20\lesssim z\lesssim 300$) is expected to broaden the observationally accessible range of scales of primordial perturbations by several decades in $k$, dramatically tightening constraints on possible large-scale behavior of $P(k)$. Improved control of systematics, nonlinearities, and foregrounds will be essential, but even partial progress may substantially narrow the allowed parameter space of the logarithmically spectral running or any other alternative soft scale dependence on very large scales.

Since the logarithmically running spectrum reproduces all current cosmological observables with the same number of free parameters as the standard power-law parameterization, present data do not statistically favor either description. However, the proposed  
Small Patch Hypothesis avoids introducing a speculative high-energy inflationary sector and remains confined within the regime of well-tested low-energy physics. From a Bayesian perspective, this corresponds to assigning a higher prior simplicity weight to the present framework than to models requiring GUT-scale scalar fields and finely tuned potentials. Unfortunately, the precise numerical value of such a Bayesian credibility prior is inherently subjective and cannot be derived from existing data alone.

In summary, the scenario explored here suggests that our observable Universe may be a small, low-entropy gravitationally smooth pocket embedded within a far larger spacetime, where the Cosmological Principle applies only locally, the Arrow of Time is an emergent phenomenon, and spatial flatness and causal connectedness (as well as the gaussianity and adiabaticity of linear perturbations) naturally follow. On sufficiently large ultra-Hubble scales, gravitational entropy density is expected to be much higher, spatial curvature need not be small, causal connectedness breaks down, and the Arrow of Time ultimately disappears in the sense that there is no preferred global time orientation associated with a saturated coarse-grained entropy bound. The proposed picture does not require primordial inflation; instead, it relies on the interplay between a very weak (e.g., logarithmic) spectral running of the primordial perturbations and the finite future de-Sitter horizon set by $\Lambda$. From this perspective, it becomes natural --- not surprising --- that we observe a Universe in which the fractional contribution of the cosmological constant contribution to the cosmic energy budget, $\Omega_{\Lambda}$, is non-negligible; the cosmological constant defines the observationally accessible region of spacetime, and large-scale homogeneity and isotropy emerge as consequences of limited observational access rather than uniquely inflationary initial conditions.

\vspace{6pt}

\funding{This research was supported by a grant from the Joan and Irwin Jacobs donor-advised fund at the JCF (San Diego, CA, USA).}

\dataavailability{The original contributions presented in the study are included in the article. Further inquiries can be directed to the authors.}

\acknowledgments{The author thanks the anonymous referees for constructive comments that
helped improve the clarity and presentation of this work.}

\conflictsofinterest{The author declares no conflicts of interest. The funders had no role in the design of the study; in the collection, analyses, or interpretation of the data; in the writing of the manuscript; or in the decision to publish the results.}

\appendixtitles{yes} 
\appendixstart
\appendix
\section[\appendixname~\thesection]{Confronting the Logarithmically Running Spectrum with~Observations}\label{appA}

This Appendix describes in some details the model selection procedure 
employed in this work that culminated in the conclusion that the spectrally running power spectrum 
considered here is statistically indistinguishable from the conventional red-tilted power spectrum from 
a global parameter estimation perspective. From a Bayesian standpoint, both power spectra 
result in very similar volumes of the posterior distributions in the multi-dimensional parameter space.
For model selection data is used from the Planck~\cite{Planck2020} CMB satellite, in addition to 
large scale structure data from the Dark Energy Survey (DES), Baryonic Oscillations (BAO), 
Type Ia supernovae (SN1a), and local measurements of $H_{0}$.

Before providing more details about how this model comparison is actually carried out it should be stressed once more that the validity of the proposed Small Patch Hypothesis does not depend on the specific choice of a logarithmic spectral dependence of the power spectrum proposed in Equation~(2). This spectrum was put forward only as a concrete illustrative example for a finite size hosting Universe, which is much larger than the observable Universe still. The parent Universe could be equally well infinite in extension with the standard red-tilted power law instead of the logarithmic power spectrum considered here.

As in the case of a power law, the pivot scale $k_{0}$ in the proposed logarithmically running form of $P_{s}(k)$ carries no physical significance. Starting from $P_{s}(k)=\frac{A_{s}}{1-(n_{s}-1)\ln(k/k_{0})}$, a redefinition $k_{0}\rightarrow k_{\star}$ can be effected such that an equivalent parameterization is obtained,
\begin{eqnarray}
P_{s}(k)=\frac{A'_{s}}{1-(n'_{s}-1)\ln(k/k_{\star})},
\end{eqnarray}
with $A'_{s}\equiv\frac{A_{s}}{1+(n_{s}-1)\ln(k_{\star}/k_{0})}$ and $n'_{s}-1\equiv\frac{n_{s}-1}{1+(n_{s}-1)\ln(k_{\star}/k_{0})}$. 

Here we adopt the standard pivot choice $k_{0}=0.05~{\rm Mpc^{-1}}$ and analyze ten dataset combinations; five include the SH0ES prior on the locally measured value of $H_{0}$, and five do not. The combinations used are; Planck alone, Planck+DES, Planck+BAO, Planck+Pantheon, and Planck+Pantheon+BAO, with and without the SH0ES prior. 
All datasets are available in the CosmoMC 2021 release.
These include Planck 2018 CMB temperature and polarization data, DES Year-1 clustering and weak lensing~\cite{DES2021}, BAO measurements from BOSS DR12~\cite{SDSS2017}, MGS~\cite{SDSSDR72015}, and 6dF~\cite{6dF2011}, and the Pantheon SNIa compilation~\cite{Pantheon2018}. The Planck data probe multipoles $2<\ell<2500$. Sampling from the posterior distributions in our analysis employs standard Gelman–Rubin convergence criteria requiring $R-1<0.01$.

\textls[-25]{To evaluate model preference the Deviance Information Criterion (DIC)~\cite{Liddle2007} is computed,}
\begin{eqnarray}
{\rm DIC}\equiv 2\overline{\chi^{2}(\theta)}-\chi^{2}(\overline{\theta}),
\end{eqnarray}
with $\theta$ denoting the model parameter vector and overbars denoting posterior expectation values. A standard parameterization of the cosmological model is employed with fundamental parameters $\Omega_{b}h^{2}$, $\Omega_{c}h^{2}$, $\theta_{MC}$, $A_{s}$, $n_{s}$ and $\tau$, so that the parameter space is six-dimensional. 
Here, $\Omega_{b}$ and $\Omega_{c}$ are the energy densities of baryons and cold dark matter in critical density units, $h\equiv H_{0}/100$ 
where $H_{0}$ is the (derived) Hubble parameter in standard km/sec/Mpc units, $\theta_{MC}$ is the typical angular scale of CMB anisotropy, and $\tau$ is the optical depth to reionization. As usual, $A_{s}$ and $n_{s}$ gauge the amplitude and spectral tilt of $P(k)$. The flat priors imposed on these fundamental parameters are the default CosmoMC ranges.

According to the Jeffreys interpretive scale~\cite{Liddle2007}, DIC differences $<$1, $1$--$2.5$, $2.5$--$5$, and $>$5 correspond to inconclusive, weak/moderate, moderate/strong, and decisive model preference, respectively. For every dataset combination analyzed here, the DIC difference between the standard power-law model and the logarithmically running spectrum remains below unity, $\Delta$DIC$\lesssim 0.6$, indicating no statistically meaningful preference for either model given current data.

Future observations, particularly high-resolution 21-cm tomography of the Dark Ages ($30\lesssim z\lesssim 200$), are expected {to} probe $P(k)$ down to kpc scales, thereby extending the leverage on scales by three decades. If this nominal potential is fully realized, such measurements could potentially distinguish between the two parameterizations. Forecast analyses using Fisher or full Bayesian evidence techniques (e.g.,~\cite{Heavens2007,Heavens2009}) already suggest that forthcoming 21-cm cosmology may have sufficient statistical leverage to meaningfully discriminate various spectral dependencies of $P(k)$, although systematics are expected to pose significant technical challenges~\cite{Shimon2013}.

In summary, current cosmological observations are fully consistent with both the standard power-law spectrum and the logarithmically running alternative considered in the present work. Currently, the two models are empirically indistinguishable. Nonetheless, if future observations were to favor a logarithmic form (or any other monotonically running alternative form for that matter), it would be consistent with the broader possibility explored here --- that the observable Universe is merely a small patch embedded within a vastly larger spacetime, and less so with cosmic inflation.

\begin{adjustwidth}{-\extralength}{0cm}
\reftitle{References}

\PublishersNote{}
\end{adjustwidth}

\end{document}